\documentclass[12pt]{iopart}

\newcommand{\up}{{\mid \uparrow \rangle}}
\newcommand{\down}{{\mid \downarrow \rangle}}

\newcommand{\bbar}{{\mid \uparrow, 7/2 \rangle}}
\newcommand{\abar}{{\mid \downarrow, 7/2 \rangle}}
\renewcommand{\a}{{\mid \uparrow, -7/2 \rangle}}
\renewcommand{\b}{{\mid \downarrow, -7/2 \rangle}}

\newcommand{\ex}{{\mid \Gamma_2^l \rangle}}
\newcommand{\exa}{{\mid \Gamma_2^l, -7/2 \rangle}}
\newcommand{\exabar}{{\mid \Gamma_2^l, 7/2 \rangle}}
\newcommand{\LH}{{{\rm LiHoF_4}}}
\newcommand{\LHx}{{{\rm LiHo_xY_{1-x}F_4}}}
\newcommand{\Ht}{{H_{\perp}}}

\newcommand{\mub}{{\mu_{\rm B}}}
\newcommand{\psio}{{\mid \psi_o \rangle}}

\newcommand{\bpsio}{{\langle \psi_o \mid}}

\usepackage{graphicx}
\usepackage{color,psfrag,epsfig}
\begin{document}

\title[]{Quantum spin glass in anisotropic dipolar systems}

\author{M. Schechter,$^1$, P. C. E. Stamp,$^{1,2}$ and N. Laflorencie$^1$}

\address{1. Department of Physics and Astronomy,
University of British Columbia, Vancouver, BC, Canada V6T 1Z1 \\
2. Pacific Institute of Theoretical Physics, University of British
Columbia, Vancouver, BC, Canada V6T 1Z1}
 \ead{moshesc@phas.ubc.ca}
\begin{abstract}
The spin-glass phase in the $\LHx$ compound is considered. At zero
transverse field this system is well described by the classical
Ising model. At finite transverse field deviations from the
transverse field quantum Ising model are significant, and one must
take properly into account the hyperfine interactions, the
off-diagonal terms in the dipolar interactions, and details of the
full $J=8$ spin Hamiltonian to obtain the correct physical picture.
In particular, the system is not a spin glass at finite transverse
fields and does not show quantum criticality.

\end{abstract}



\section{Introduction}
The study of spin glasses, and in particular the classical
transition between the spin-glass (SG) to the paramagnetic (PM)
phase, have been thoroughly studied since the $70$'s \cite{BY86}. As
interest in quantum phase transitions has grown, the understanding
of quantum criticality at the SG to PM phase transition has drawn
much theoretical interest \cite{MH93,You97,Sac99}. Experimentally,
the quantum phase transition is not accessible in most spin glasses,
where the strong exchange interaction blocks quantum fluctuations at
available magnetic fields. However, anisotropic magnetic dipolar
systems, notably the $\LHx$ compound, have very weak exchange
interactions, and seem ideal for the observation of such a
transition \cite{Ros96}. In this system the dipolar interaction and
single-ion anisotropy terms have magnitude $\sim 1$K and $\sim 10$K
respectively; with spin $J=8$, appreciable quantum fluctuations are
expected already at transverse fields $\Ht \sim 1$T. The strong easy
axis anisotropy means that for $T \ll 10K$, the Ho spins truncate to
an Ising-like doublet; this anisotropy also strongly suppresses all
but the longitudinal terms of the dipolar interaction. The $\LH$
compound can be diluted by exchanging Ho with the nonmagnetic Y
atom, resulting in the $\LHx$ compound with any desired x
\cite{Ros96}. The angular dependence of the dipolar interaction, in
the presence of quenched randomness, results in frustration. Thus,
the system which is ferromagnetic at x$=1$ \cite{BRA96}, turns into
a SG for x$=0.167$, with $T_c=0.13$K \cite{WER+91,WBRA93}. For the
latter dilution, the SG to PM transition was studied as a function
of $T$ and $\Ht$\cite{WER+91,WBRA93}. Both the linear and nonlinear
susceptibility were measured. Despite the importance of this
experiment, some of its very interesting features long remained
without a proper explanation. Questions included (i) the reduction
of the cusp in the nonlinear susceptibility with {\it decreasing}
$T$; (ii) the observation that it is much easier to disorder the SG
thermally than quantum mechanically; (iii) the sharpness of the low
$T$ crossover between a PM response and slow relaxation; and (iv)
the smallness of the low $T$ critical exponent, as observed in the
nonlinear susceptibility near the transition.

Theoretically, the $\LHx$ system was considered to be a good
realization of the transverse field Ising model (TFIM) in the
electronic degrees of freedom. Recently \cite{SS05,SL05,SS06} it was
shown that the real system differs from the above model in two
significant ways, which affect the physics considerably; (a) the
hyperfine (hf) interaction between the Ho electronic and nuclear
spins is strong, and for x$\ll 1$ in general, and in the SG
experiments \cite{WBRA93} in particular, dominate the physics
\cite{SS05,SS06} (b) the off-diagonal dipolar terms, although
effectively reduced, become essential at any finite $\Ht$ as they
reduce the symmetry of the model \cite{SL05}. In this paper we show
how these features answer the first two of the four questions noted
above; we also show that a proper understanding of the off-diagonal
dipolar terms requires going outside the simple TFIM.
\section{Hyperfine interactions}
The Hamiltonian describing the $\LHx$ system is given by a sum of
crystal field \cite{GWT+01,CHK+04}, Zeeman, hf, and inter-Ho
interaction terms:
\begin{equation}
H = H_{\rm cf} + H_{\rm Z} + H_{\rm hf} + H_{\rm int} \, .
\label{generalH}
\end{equation}
The Ho ion has a $J=8$ angular momentum. $H_{\rm cf}$ splits the
$17$-fold degeneracy, leaving 3 relevant low energy levels - an
Ising-like doublet, denoted $\up, \down$, and a first excited state
approximately $10$K higher in energy, denoted $\ex$
\cite{GWT+01,SS05}. $H_Z=- \sum_i g_J \mub \vec{H} \cdot \vec{J_i}$,
is the Zeeman energy, and $H_{int} = - \sum_{ij} U_{ij}^{\alpha
\beta} J_i^\alpha J_j^\beta$ is dominated by the dipolar interaction
\cite{CHK+04}. We denote the easy axes by z, and consider $\Ht \|
$x. It is common to neglect all but the longitudinal terms in the
dipolar interaction, and drop the hf interaction, thus obtaining a
low energy TFIM effective Hamiltonian:
\begin{equation}
  H =  - \sum_{i,j} V_{ij}^{zz} \tau_i^z  \tau_j^z - \Delta_0
(H_{\perp}) \sum_i \tau_i^x \, ,
 \label{IsingH}
\end{equation}
where $\vec{\tau}_j$ is a Pauli vector describing the two-level
effective electronic spin at spatial position ${\bf r} = {\bf r}_j$,
and $\Delta_0 \propto \Ht^2/\Omega_0$ for small $\Ht$. However, both
the hf and off-diagonal dipolar interactions are of crucial
importance for the $\LHx$ system.
\begin{figure}
\begin{center}
\includegraphics[width = 0.5\columnwidth]{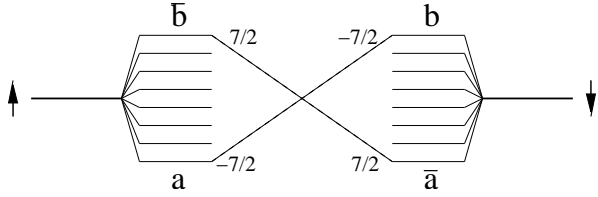}
 \caption{Splitting of the electronic low energy doublet ($\uparrow$ and
$\downarrow$) due to the longitudinal hf interaction. The GS
doublet, $a$ and $\bar{a}$ have a definite and opposite nuclear
spin, $\pm 7/2$. Transverse magnetic field couples states with the
same nuclear spin, as is shown by the dashed lines. }
\end{center}
    \label{fignuclearsplitting}
\end{figure}
The Ho atom is a pure isotope $I=7/2$ nuclear spin with contact hf
interaction $H_{\rm hf} = A_J \sum_i \vec{I_i} \cdot \vec{J_i}$. Due
to the strong anisotropy we consider first the longitudinal part of
the hf interaction $H_{\rm hf}^{\|} = A_J I^z J^z$. This term splits
each of the states $\up,\down$ into an eightfold multiplet of nearly
equidistant levels, with separation $\sim 205 mK$ \cite{GWT+01}
between adjacent levels (Fig. 1).  This splitting, larger than the
typical dipolar energy and the relevant experimental temperatures
\cite{WBRA93}, influences significantly the physics of the system.
The Ising doublet states have now a definite nuclear spin, i.e.
$I_z=-7/2$ for the electronic state $\up$ and $I_z=7/2$ for the
electronic state $\down$. A transverse field $\Ht$ then couples $a
\equiv \a$ with $b \equiv \b$ and $\bar{a} \equiv \abar$ with
$\bar{b} \equiv \bbar$ (see Fig. 1). Thus, the transverse field by
itself does not induce quantum fluctuations between the relevant
Ising doublet GSs, but only re-normalizes their effective spin. The
{\it transverse} hf interactions $H_{\rm hf}^{\perp}=A_J [I^z J^z +
(I^+ J^- + I^- J^+)/2]$ allow simultaneous changes in $I_z$ and
$J_z$, and thence quantum fluctuations between the Ising doublet
states. However, as is detailed in Refs. \cite{SS05,SS06} this
mechanism is negligible for $H_{\perp} \ll \Omega_0/\mub$. Thus, the
inclusion of the hf interactions results in three energy scales
which dictate the behavior of the system, and in particular the
position of the crossover between the SG and PM phases as function
of $T$ and $\Ht$ \cite{SS05,SS06}. At zero field, $T_c(0)$ is
dictated by the dipolar interaction $V_0$, since the nature of the
Ising doublet is not important for the classical transition. The
behaviour of $T_c(\Ht)$ with $\Ht$ is governed by the strength of
the hf interaction and the crossover at $T=0$ is governed by the
anisotropy energy $\Omega_0$, as quantum fluctuations between the
relevant Ising states become significant only when the state $\ex$
becomes appreciably hybridized with $\up, \down$. All three energy
scales become apparent in the position of the line separating the SG
and PM phases, and in particular the relation $V_0 \ll \Omega_0$
explains the fact that it is much easier to disorder the SG
thermally rather than quantum mechanically \cite{SS05,SS06}.

\section{Off-diagonal terms of the dipolar interaction}
The longitudinal term of the dipolar interactions, $\propto J_i^z
J_j^z$, has direct matrix elements within the low energy Ising
states $\a, \abar$. All other interaction terms involve the state
$\ex$ in second or higher perturbation expansion. Since $\Omega_0
\gg V_0$, one is tempted to neglect all but the longitudinal
interaction. However, the off-diagonal terms, and in particular the
terms $\propto J_i^z J_j^x$ become important at $\Ht \neq 0$, as
they change the symmetry of the system. The $J_z \rightarrow -J_z$
symmetry, while maintained by $\Ht$ without the off-diagonal terms,
is destroyed. This reduction of symmetry results in a generation of
an effective random field at each site \cite{SL05}. As a result,
within the droplet picture of Fisher and Huse \cite{FH86-88} and
using an Imry-Ma like argument \cite{IM75} one obtains \cite{SL05} a
magnetic field dependent finite correlation length. The system with
{\it transverse} magnetic field and off-diagonal dipolar
interactions becomes equivalent to the random field Ising model,
i.e. a SG in the presence of random {\it longitudinal} magnetic
field. Thus, the scaling theory of Fisher and Huse \cite{FH86-88}
predicts an instability of the SG phase to finite $\Ht$ in our case,
equivalent to its prediction of the absence of de-Almeida Thouless
line in the RFIM. Interestingly, to obtain the correct physical
picture of the system at finite field one has to consider the large
Ho spin, going beyond the simplified Ising picture. The reason is
that it is the fluctuations between each of the single Ho GSs and
its first excited states, and not the much smaller quantum
fluctuations between the two Ising states $a, \bar{a}$ \cite{SL05},
that govern the magnitude of the effective random field and the
reduction of the correlation length.

Within the scaling picture of Fisher and Huse \cite{FH86-88} the SG
at zero transverse field has two time-reversed GSs, denoted $\psi$
and $\bar{\psi}$. Each Ho ion in the GS $\psi$ is in either states
$a$ or $\bar{a}$, and in the opposite state in the GS $\bar{\psi}$.
Consider a single Ho ion at $\Ht \neq 0$. For small magnetic fields
the fluctuations between the Ising states $a$ and $\bar{a}$ are
negligible \cite{SS05}. Yet, the energy of each of the Ising states
is reduced by an energy proportional to $\Ht^2/\Omega_0$, due to
fluctuations to the relevant excited states ($\exa$ for $a$ and
$\exabar$ for $\bar{a}$). If we choose an arbitrary region in state
$\psi$, the energy reduction due to the field is just the sum over
all spins of the single spin energy gain. With $J_z \rightarrow
-J_z$ symmetry, the same energy reduction occurs for the state
$\bar{\psi}$. Now consider the effect of the off-diagonal dipolar
terms (in particular, the term $V_{ij}^{zx} J_i^z  J_j^x$) on the
different domains. In second order perturbation, the domain energy
shift is given by \cite{SL05}

\begin{equation}
 E_{\psi}^{(2)} =  - \frac{\bpsio ( \sum_{i \neq j} V_{ij}^{zx}
J_i^z  J_j^x + \mub \Ht \sum_i J_i^x)^2 \psio}{\Omega_0} \, .
 \label{E2}
\end{equation}
The dipolar terms have randomly the same or the opposite sign to
that of the magnetic field, and typically, by flipping a domain of
$N$ spins one gains an energy of \cite{SL05}

\begin{equation}
\langle \delta E \rangle= c \frac{j^2 \mub \Ht V_0
\sqrt{N}}{\Omega_0} \, ,
 \label{deq}
\end{equation}
where $j=\max{J_z}$. Comparing this energy gain to the energy cost
of flipping the domain\cite{SL05}, $\approx j^2 V_0 L^{\theta_d}$
where $L$ is the domain linear size, one finds a finite correlation
length at any $\Ht$ given by\cite{SL05}

\begin{equation}
\xi \approx \left(\frac{\Omega_0}{\mub \Ht}\right)^\frac{1}{(3/2) -
\theta_d} \, .
 \label{corl}
\end{equation}
Importantly, this correlation length depends only on $\Ht$ and
$\Omega_0$. In the experiment \cite{WBRA93}, as $T$ is decreased the
crossover to the PM phase occurs at higher $\Ht$, dictating a
smaller correlation length and a reduced cusp in the nonlinear
susceptibility \cite{SL05}. Note that the cusp is further reduced
due to the renormalization of the effective spin \cite{SS05}.
Interestingly, the finite correlation length results in an enhanced
transverse field in the x direction\cite{SL05}, which was
anticipated by the comparison of the experimental and theoretical
positions of the crossover line between the SG and PM
phases\cite{SS05}.

In principle, one could also get a finite correlation length in a
spin-half Ising model, by introducing a longitudinal interaction
$\propto V_0$ and a reduced off-diagonal interaction $\propto \alpha
V_0$\cite{SL05}. Similar considerations to the ones above lead to a
correlation length $\xi \approx [V_o/(\alpha \mub
\Ht)]^{1/(3/2-\theta_d)}$ \cite{SL05}, coming from quantum
fluctuations between the Ising doublet states, with a $\xi$
depending on $V_0$. A notable difference between this result and the
correct one is the size $\xi$ at the crossover to the PM phase at
$T=0$. This crossover actually occurs at $\Ht \approx \Omega_0$ for
the case of large spin \cite{SS05,SL05}, leading to $\xi \approx 1$.
However in the effective spin-half model the crossover occurs at
$\Ht \approx V_0$ and $\xi \approx 1/\alpha$ at the crossover,
emphasizing the inadequacy of the spin-$\frac{1}{2}$ model.
Importantly, our results here are easily generalized to any
anisotropic systems, as long as a dipolar interaction exists
\cite{SL05}.

\section{Numerical results}
\begin{figure}[!ht]
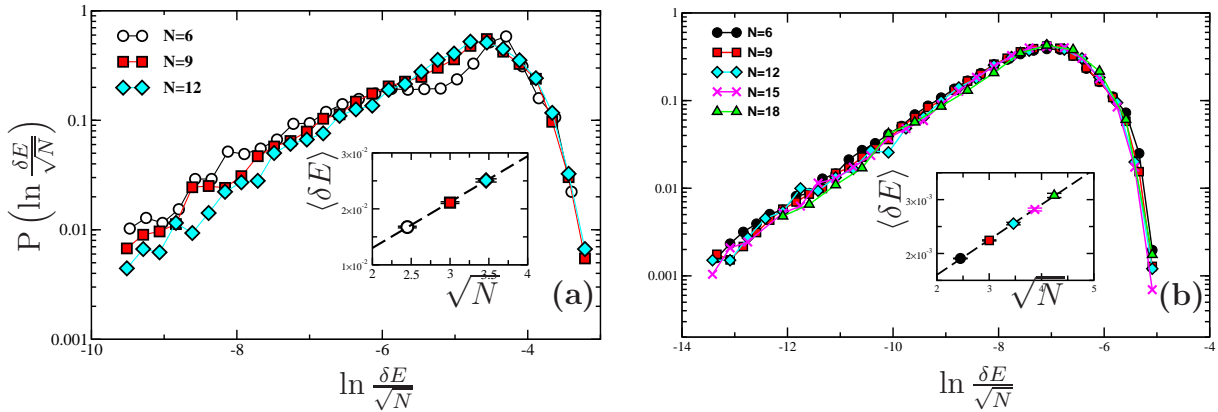

\begin{minipage}{0.5\textwidth}
\psfrag{L}{$\ln \frac{\delta E}{\sqrt{N}}$}
\psfrag{P}{${\rm{P}}\left(\ln\frac{\delta E}{\sqrt{N}}\right)$}
\psfrag{B}{${\langle{\delta E\rangle}}$}
\psfrag{A}{$\sqrt{N}~~~~$ {\bf{(a)}}}
\begin{center}
\epsfig{file=GapSpinOneOmega2Delta0.02.eps,width=7.7cm,clip}
\end{center}
\end{minipage}
\begin{minipage}{0.5\textwidth}
\psfrag{L}{$\ln \frac{\delta E}{\sqrt{N}}$}
\psfrag{P}{$~$}
\psfrag{B}{${\langle{\delta E\rangle}}$}
\psfrag{A}{$\sqrt{N}~~~~~~~~$ {\bf{(b)}}}
\begin{center}
\epsfig{file=GapSpinHalfJp0.1Delta0.0002.eps,width=8cm,clip}
\end{center}
\end{minipage}
\caption{Distributions of the finite size gaps $\delta E$ rescaled by
  $\sqrt{N}$ and plotted in a semi-log scale.
  Lanczos ED data collected over 10,000 random diluted $\LHx$ samples
  for each size $N$. Insets: Linear
  dependence of the disorder average gap $\langle \delta E\rangle$
  vs $\sqrt{N}$.
(a) Results obtained for the spin-1 Hamiltonian
  (\ref{s1}) with $\Omega_0/\mub H_t=100$ for three different sizes, with $x=18.75\%$.
  (b) Results obtained for the spin-$\frac{1}{2}$ model
    (\ref{halfH}) with $\alpha=0.1$, $\mub H_t=0.0025$ and five different sizes with a dilution x$=1/12$.}
\label{fig:gaps}
\end{figure}

The scaling relation (\ref{deq}) has been checked against Lanczos
exact diagonalization (ED) computations on finite size
clusters~\cite{SL05}. In order to get closer to the experiment, we
randomly distribute $N$ moments at the rare earth sites of three
dimensional $\LHx$ diluted lattices, and focus on $s=1$ particles
with an on-site anisotropy $\Omega_0\simeq 10$K which accounts for
the crystal field. Therefore, the following transverse field dipolar
spin-1 Hamiltonian
\begin{equation}
{\cal{H}}_{1} = - \sum_{i \neq j}\left[\frac{1}{2}V_{ij}^{zz}S_i^z
S_j^z +V_{ij}^{zx}S_i^z  S_j^x\right] -\mub H_t\sum_iS_{i}^{x} -
\Omega_0 \sum_i\left([S_i^z]^2 - s^2\right) \, \label{s1}
\end{equation}
has been diagonalized on $\LHx$ lattices with $x=18.75\%$ for
various sizes, and over 10,000 independent random samples for each size.
In the perturbative regime, we have computed the finite size gap $\delta E$ for
each sample and the $\sqrt{N}$ scaling stated in Eq. (\ref{deq}) is clearly
demonstrated, as shown in Fig. \ref{fig:gaps} (a).

A similar calculation was done for the spin-$\frac{1}{2}$ Ising
model
\begin{equation}
  {\cal{H}}_{\frac{1}{2}} =  - \sum_{i\neq j}\left[ \frac{1}{2}V_{ij}^{zz}
S_i^z  S_j^z+\alpha V_{ij}^{xz} S_i^x  S_j^z \right]- \mub H_t
\sum_i S_i^x. \label{halfH}
\end{equation}
Numerically speaking, the $s=1/2$ problem is easier since it leads
to a smaller Hilbert space dimension and allows us to check the
$\sqrt{N}$ scaling relation over a broader range of sample sizes, as
shown in Fig. \ref{fig:gaps} (b). However, here too the essential
difference between the spin-$\frac{1}{2}$ and spin-$1$ models is
apparent. In the spin-$\frac{1}{2}$ model, since the significant
fluctuations are between the Ising doublet states, the strength of
the dipolar interaction replaces $\Omega_0$ in the denominator of
Eq. (\ref{E2}), which results in better convergence of the numerics
at the smallest sizes.

\section{Conclusion}
Of the four experimental puzzles mentioned in the introduction, the
first two are explained by our analysis. Since we believe that the
Hamiltonian (\ref{generalH}) includes all the physics relevant to
the experiment \cite{WBRA93}, we expect that the solution to the
latter two puzzles lies within the same framework of considerations
above.

\end{document}